\documentclass{osa-article}
\journal{osac}

\articletype{Research Article}

\begin{document}

\title{Generation of a microresonator soliton comb via current modulation of a DFB laser}

\author{KENJI NISHIMOTO,\authormark{1} KAORU MINOSHIMA,\authormark{2,3} TAKESHI YASUI,\authormark{1,3} and NAOYA KUSE\authormark{3,4,*}}

\address{\authormark{1}Graduate school of Technology, Industrial and Social Science, Tokushima University, 2-1, Minami-josanjima, Tokushima, Tokushima 770-8506, Japan\\
\authormark{2}Graduate School of Informatics and Engineering, The University of Electro-Communications, 1-5-1, Chofugaoka, Chofu, Tokyo 182-8585, Japan\\
\authormark{3}Institute of Post-LED Photonics, Tokushima University, 2-1, Minami-Josanjima, Tokushima, Tokushima 770-8506, Japan\\
\authormark{4}PRESTO, Japan Science and Technology Agency, 4-1-8 Honcho, Kawaguchi, Saitama, 332-0012, Japan}

\email{\authormark{*}kuse.naoya@tokushima-u.ac.jp} 


\begin{abstract}
Dissipative Kerr-microresonator soliton combs (hereafter called soliton combs) has been rapidly progressing as compact frequency combs. Comb mode scanning of the soliton combs with a large range and fast speed is of paramount importance for applications such as LiDAR and spectroscopy, requiring large and rapid frequency scanning of a pump continuous-wave (CW) laser as well as resonance frequency of a microresonator. Here, we demonstrate the generation of a soliton comb by a distributed feedback (DFB) laser toward the comb mode scanning with a large range and fast speed. Compared with conventional pump CW lasers (i.e. external cavity diode lasers: ECDLs), DFB lasers can be frequency-scanned more largely and rapidly without mode-hopping. In addition, because of the fast scan speed of the DFB laser, a single soliton comb is generated simply by controlling the injection current of the DFB laser, greatly simplifying the system without having any additional optical modulators such as a carrier-suppressed single-sideband modulator (CS-SSB modulator), acousto-optic modulator (AOM), and auxiliary CW laser.
\end{abstract}

\section{Introduction}
Optical frequency combs \cite{diddams2020optical} have revolutionized in the fields of metrology, including optical atomic clocks \cite{oelker2019demonstration, nemitz2016frequency, stern2020direct}
, distance measurement \cite{minoshima2000high, ye2004absolute,coddington2009rapid, na2020ultrafast, riemensberger2020massively}, microwave generation \cite{fortier2011generation, nakamura2020coherent, liu2020photonic}, to name a few. Since the discovery of dissipative Kerr soliton combs (hereafter called soliton combs) \cite{Herr_soliton, kippenberg2018dissipative} in high Q microresonators as potentially, fully chip-scale optical frequency combs \cite{stern2018battery,shen2020integrated}, the soliton combs have been attracting significant attention. Soliton comb is a mode-locked state with high coherence and ultra-short pulse trains. The comb spacing of soliton combs can be more than 100 GHz because of the very short cavity length, which provides unique applications beyond the use in research laboratories such as ranging (including light detection and ranging, LiDAR) \cite{trocha2018ultrafast,riemensberger2020massively}, coherent optical telecommunication \cite{marin2017microresonator}, and mmW/THz waves generation \cite{zhang2019terahertz}.

Continuous scanning of the comb modes of soliton combs further provides attractive applications of the soliton combs such as massively parallel frequency-modulated (FM) CW LiDAR \cite{riemensberger2020massively}, high resolution LiDAR based on FMcomb LiDAR \cite{kuse2019frequency}, and broadband, high-frequency resolution spectroscopy \cite{yu2017microresonator, kuse2020continuous, lin2020broadband}. In the applications, scan range and speed are important, because a large scan range enables high depth resolution, while fast scanning enables short measurement time in the LiDAR applications. In spectroscopy, large scanning eliminates blind frequency areas caused by the sparsity of the comb modes, also enabling high-frequency resolution determined by the linewidth of the comb modes, not by the comb mode spacing. Besides, fast scanning allows fast measurement time. 

To largely scan the comb modes (\textgreater 100 GHz), both a pump CW laser and the resonance frequency of a microresonator have to be scanned simultaneously \cite{riemensberger2020massively,liu2020monolithic}, because the soliton exists only when the pump CW laser is red-detuned from the resonance frequency by a few GHz. A microheater deposited on a microresonator has been utilized for the large range of the comb mode scanning \cite{xue2016thermal,Gaeta_heater}. By implementing a feedback loop based on Pound-Drever-Hall (PDH) locking or soliton power locking to fix the detuning while the comb modes are scanned, comb mode scanning of 190 GHz \cite{kuse2020continuous} and 31 GHz \cite{lin2020broadband} has been demonstrated. In the demonstrations, a pump CW laser based on an external cavity diode laser (ECDL) has been used, which impedes the simultaneous realization of large scan range and fast scan. Scan range and speed of the comb modes of a soliton comb from an ECDL would be at best around 30 GHz and 1 ms limited by a PZT inside the ECDL, respectively (note that larger scan range is possible if slower scan speed is allowed). Although a compact semiconductor laser with an external fiber Bragg grating (FBG) has been also used for the generation of a soliton comb, the scan speed of the laser cannot be fast (> 1 ms) due to the slow response of the FBG \cite{raja2020chip}. Alternatively, distributed feedback (DFB) lasers are more promising for the large scan range and fast scanning because the frequency of DFB lasers can be largely (\textgreater 100 GHz) and rapidly (\textless	 1 ms) scanned without any mode-hopping. Very recently, soliton combs have been generated from DFB lasers, in which self-injection locking (SIL) is utilized. In SIL \cite{stern2018battery, voloshin2019dynamics,raja2019electrically,shen2020integrated}, a DFB laser is butt-coupled to a high Q Si$_3$N$_4$ (SiN)-based microresonator (\textgreater 10\textsuperscript{7}), and Rayleigh back-scattering from the microresonator is injected into the DFB laser, enforcing the oscillation frequency of the pump CW laser to be equal (or with an offset in some cases) to the resonance frequency of the microresonator when the resonance frequency is within the injection locking range of the DFB laser. However, it is not clear for the DFB laser with SIL to allow the comb mode scanning of a large range and fast speed because the detuning between the frequency of the injected DFB laser and a resonance frequency of the high Q microresonator is inherently determined by the optical phase of the back-scattered light and optical power coupled to the microresonator, which also makes it difficult to access a single soliton comb. Therefore, for the comb mode scanning, independent control of the resonance frequency and pump CW laser might be more convenient. However, the soliton comb from a DFB laser without the use of SIL has not been demonstrated to the best of our knowledge. 

In this paper, we demonstrate the generation of a single soliton comb from a microresonator pumped with a DFB laser without implementing SIL. Since SIL is not used, the system is suitable for large and fast comb mode scanning. On the other hand, the thermo-optic effect in microresonators induced when a chaotic comb transitions to a soliton comb has to be overcome. There are several methods to overcome the thermo-optic effect such as fast tuning of the pump frequency by a carrier-suppressed single-sideband modulator (CS-SSB modulator) \cite{Papp_OL_SSB18, stone2018thermal}, fast tuning of the resonance frequency by pump power modulation \cite{Kippenberg_SiN16}, and the utilization of an auxiliary CW laser \cite{niu2018repetition,zhou2019soliton,zhang2019sub, lu2019deterministic}. In our demonstration, we just modulate the injection current of the DFB laser, which is enough to access a stable single soliton comb because of the fast scan speed of the DFB laser (\textgreater 10 GHz/$\mu$s). Compared with conventional soliton comb systems generated from an external cavity diode laser (ECDL) as a pump CW laser with the additional optical modulators, our method not only simplifies the system, but also is readily applicable to comb mode scanning with a large range and fast speed.

\section{Result}
\subsection{Experimental Setup}
Figure 1(a) shows the experimental setup. A DFB laser is used for a pump CW laser with an oscillation wavelength of around 1548 nm. The output from the DFB laser is amplified by an Er-doped fiber amplifier (EDFA) with one forward and two backward pump laser diodes. The maximum output from the EDFA is approximately 500 mW. The output from the EDFA is coupled into a high-Q microresonator based on SiN through a lensed fiber. The free spectral range (FSR) and Q of the microresonator are approximately 540 GHz and 10$^6$ (Fig. 1(b)), respectively. The on-chip power is about 200 mW. The output from the microresonator is split into two outputs by a 1 $\times$ 2 optical splitter. One is used to measure the optical spectrum of the generated soliton combs by an optical spectrum analyzer (OSA). The other is directed to a notch filter (NF) to reject the residual of the DFB laser to measure the time evolution of the power of the generated comb. The output from the notch filter is monitored by an oscilloscope (OSC). The oscillation frequency of the DFB laser is controlled via changing the injection current, which is controlled by an arbitrary waveform generator (AWG). The modulation speed can be as high as 10 GHz/$\mu$s as shown in Fig. 1(c). When the injection current of the DFB laser is increased, the output power of the DFB laser increases, which can affect the resonance frequency of the microresonator through the thermo-optic effect. However, in our experiment, the coupled power to the waveguide keeps almost the same because the EDFA is operated in the saturation regime.

\begin{figure}[h!]
\centering\includegraphics[width=13.3cm]{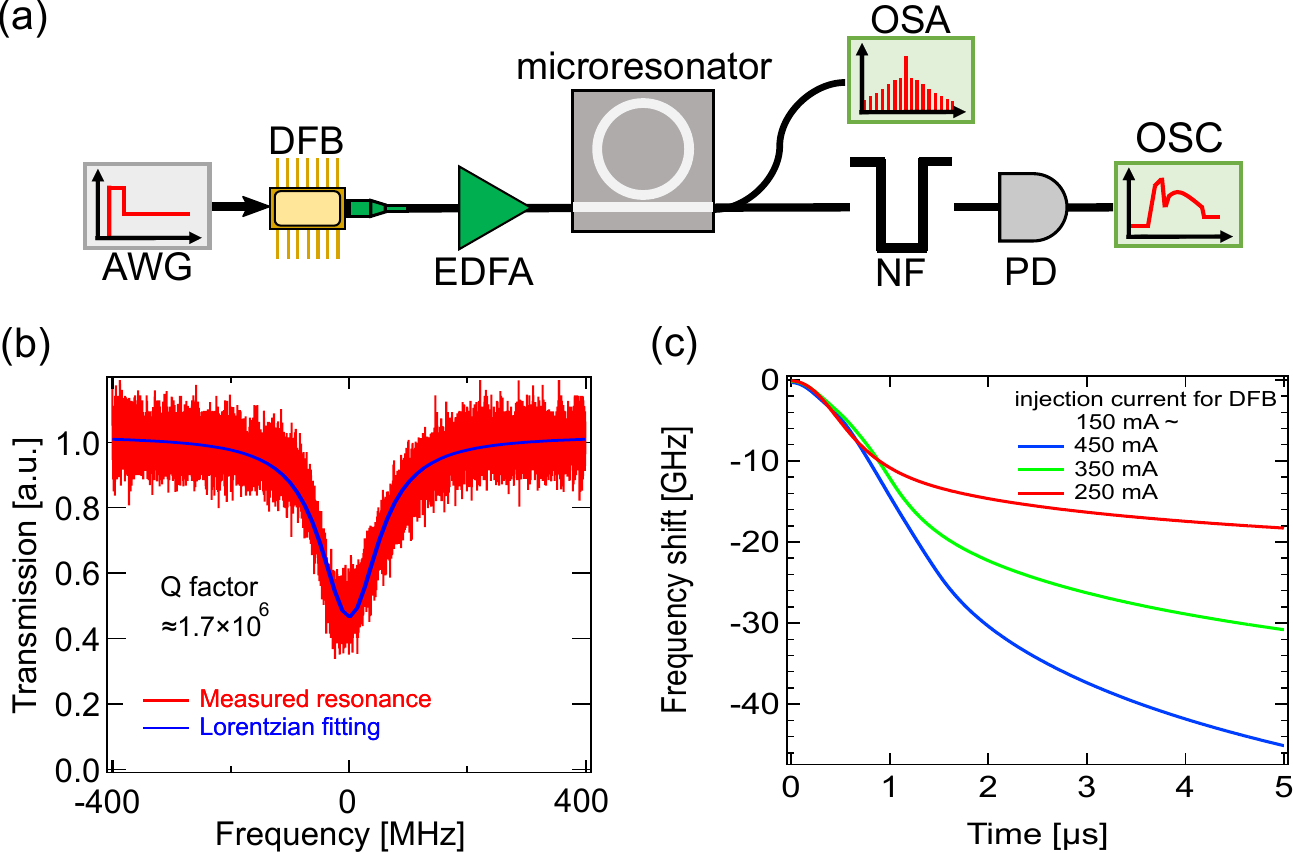}\label{setup}
\caption{(a) Schematic of the experimental setup to generate a soliton comb. (b) The red curve showed measured resonance frequency, and the blue curve shows Lorentzian fitting. (c) Frequency shift of the DFB laser, when a step function is applied, changing the injection current from 150 mA to 250 mA (red), 350 mA (green) and 450 mA (blue).}
\end{figure}

\subsection{Results of DFB laser direct modulation setup}
To access a stable soliton comb, the frequency of the DFB laser has to be controlled to follow the resonance frequency of the microresonator, since the speed of the thermo dynamics of the microresonator is comparative to the frequency scan speed of the DFB laser. A procedure to find an appropriate signal for the injection current to generate a single soliton comb is shown in Figs. 2(a) and (b). Figures 2(a) and (b) show signals to control the injection current and comb powers, respectively. In the first step, the amount of the increase in the injection current is determined. The increase of the injection current induces the decreases of the oscillation frequency of the DFB laser. When the increase of the injection current is not large enough (Fig. 2(a) - i), a chaotic comb is generated, gradually relaxing to a primary comb due to the decrease of the resonance frequency caused by the increase of the intracavity power (Fig. 2(b) - i). When the increase of the injection current is set at correctly, a soliton step is observed (Fig. 2(b) - ii).  
\begin{figure}[h!] \centering\includegraphics[width=13.3cm]{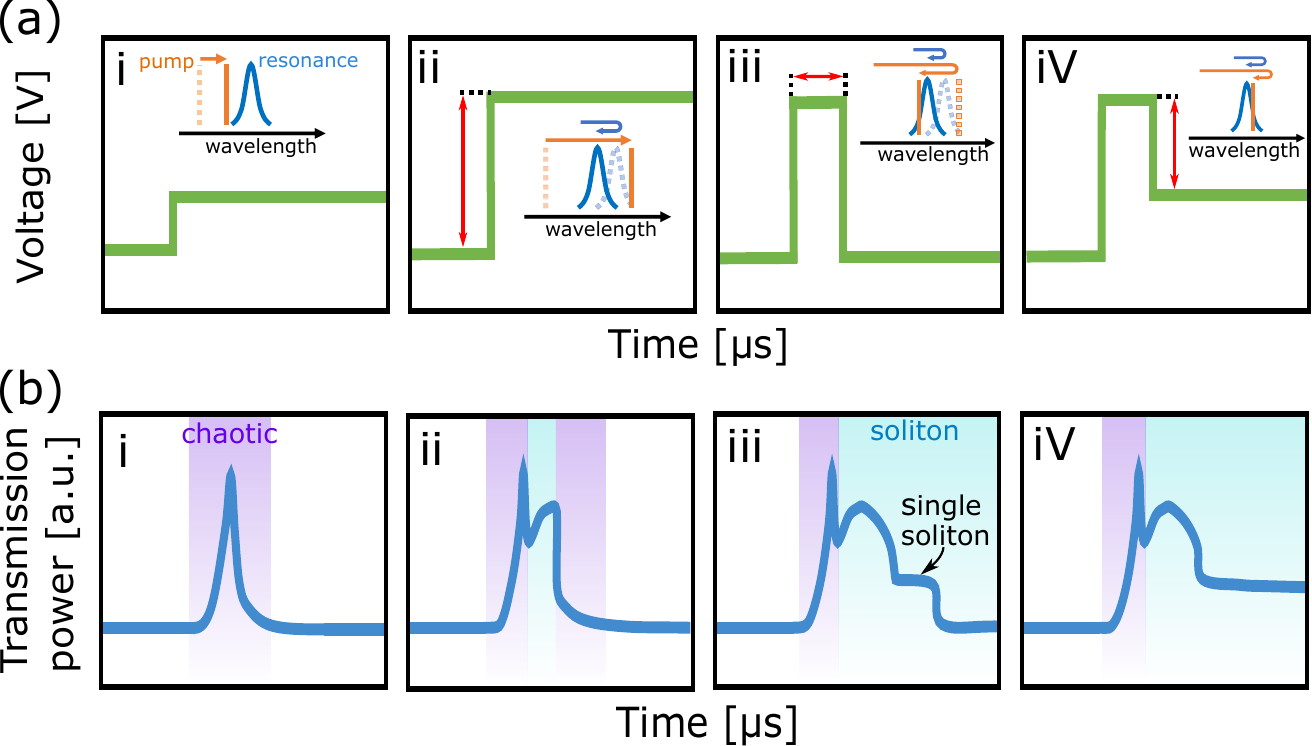}\label{solitonimage}\caption{(a) Illustrations of the control signals of the DFB laser. The insets show the relationship between the wavelength of the pump CW laser and the resonance wavelength of the microresonator. The red arrow indicates controlled parameters in each procedure. (b) Illustrations of the transmission comb power without the residual pump laser. The purple and light blue areas show the time when the chaotic comb and (multi) soliton state exist, respectively.}
\end{figure}
The increase of the injection current should not be too much but should be set at the value just above where the soliton step is observed. Although the soliton step is observed in Fig. 2(b) - ii, the soliton cannot be kept long (\textless 1 $\mu$s) because of the thermo-optic effect. The thermo-optic effect shifts the resonance frequency of the microresonator to high frequency, increasing the detuning between the frequency of the pump CW laser and the resonance frequency. In the second step, the time to decrease the injection current is determined to extend the lifetime of the soliton. In this step, the control signal to the injection current is instantaneously turned off (Fig. 2(a) - iii). Note that the frequency of the DFB laser begins to increase a little after turning off the control signal because of the delayed response of the DFB laser. When the frequency of the DFB laser begins to increase right after the transition from the chaotic to the soliton comb, the lifetime of the soliton comb is extended by mitigating the change of the detuning, followed by a single soliton state (Fig. 2(b) - iii). In the third step, the offset of the control signal is determined (Fig. 2(a) - iv), which fixes the final frequency of the DFB laser, to keep the single soliton comb for a long time (Fig. 2(b) - iv). 

The signal to generate a soliton comb used in this report is determined according to the above procedure as shown in Fig. 3(a). Note that we gradually and a little reduce the control signal between 0.1 ms and 0.4 ms to deal with the shift of the resonance frequency caused by the slow thermal effect in the microresonator as shown in the inset in Fig. 3(a). Figure. 3(b) shows the time evolution of the frequency of the DFB laser (blue) measured by an imbalanced Mach-Zehnder interferometer and the comb power (red). The decrease in the frequency of the DFB laser causes the decrease of the detuning from 0 to 1 $\mu$s. At 1.2 $\mu$s, the detuning becomes red (the frequency of the DFB laser is smaller than the resonance frequency), and the chaotic comb transitions to the soliton state. During the soliton state, the frequency of the DFB laser begins to increase, switching the comb power to decrease (the red curve in Fig. 3(b)) since the comb power is proportional to the detuning. When the detuning is further decreased, a single soliton comb is obtained. Finally, after applying the slow control signal as shown in the inset in Fig. 3(a), the detuning is fixed, and the single soliton comb is maintained for a few hours without any feedback loops. The optical spectrum of the soliton comb is shown in Fig. 3(c). The comb spacing corresponds to the FSR of the microresonator with 55 nm of 3 dB bandwidth.

\begin{figure}[h!]
\centering\includegraphics[width=13.3cm]{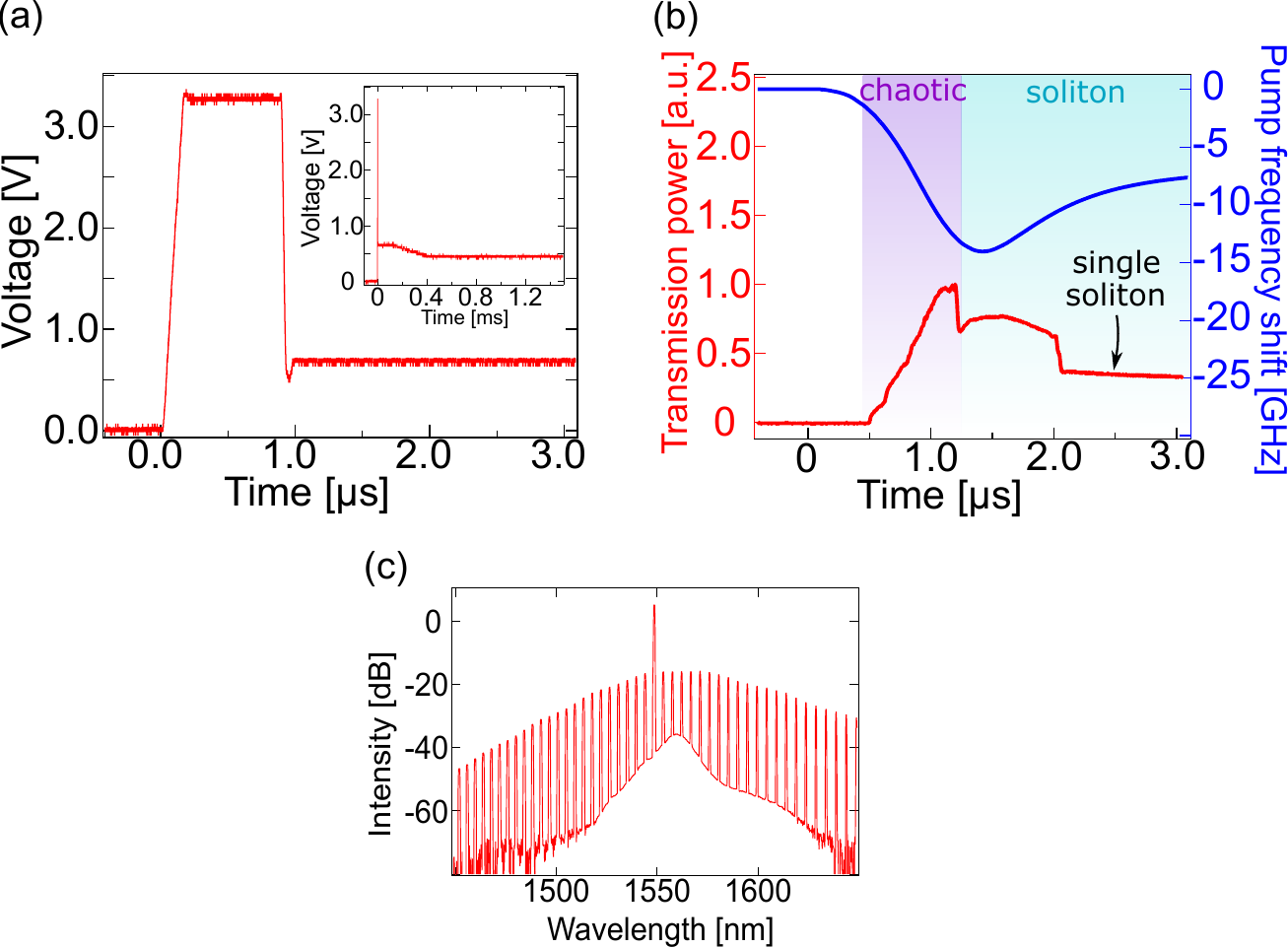}\label{combpower}
\caption{(a) The control signal for the injection current of the DFB laser. The inset shows the control signal with a short time span. (b) Frequency change of the DFB laser (blue) with the time evolution of the comb power (red). The purple and light blue area show the time when the chaotic comb and (multi) soliton state exist, respectively. (c) The optical spectrum of the soliton comb generated.}
\end{figure}

\section{Conclusion}

In conclusion, we demonstrated the generation of a soliton comb pumped by a DFB laser, in which the injection current of the DFB laser is adequately controlled. The thermo-optic effect in a microresonator, induced especially when the chaotic comb transitions to the soliton comb, is overcome by the fast scanning of the DFB laser. The demonstrated method does not require any additional optical components such as a CS-SSB modulator, AOM, and auxiliary cooling laser, enabling the simpler comb system than systems with bulky ECDLs. More importantly, since DFB lasers have a larger and faster mode-hop-free scan range than ECDLs, the soliton comb from DFB laser can be readily applied to high resolution spectroscopy \cite{kuse2020continuous} and LiDAR \cite{kuse2019frequency} along with a large range and fast speed scanning. Here, the linewidth of DFB lasers is two orders of magnitude worse than that of ECDLs, which may be a significant drawback of soliton combs generated from DFB lasers in terms of phase noise. However, in ref \cite{nishimoto2020investigation}, we showed the phase noise of the comb modes is not equal to that of pump CW lasers but is limited by the thermo-refractive noise of microresonators when an ECDL is used as a pump CW laser. Furthermore, the relative phase noise between the comb modes, which is important for mmW/THz wireless communications by soliton combs, has little difference between the soliton combs generated from a DFB laser and ECDL \cite{nishimoto2020investigation}. Therefore, we believe soliton combs generated from DFB lasers are very attractive because of the advantages; simple system, large and fast scanning.

 


\section*{Funding}
This work was financially
supported by JST PRESTO (JPMJPR1905), Japan Society for the Promotion of Science (19H00871), Cabinet Office, Government of Japan (Subsidy for Reg. Univ. and Reg. Ind. Creation), Research Foundation for Opto-Science and Technology, and Nakatani Foundation for Advancement of Measuring Technologies in Biomedical Engineering.
\\

\section*{Disclosures} The authors declare no conflicts of interest.

\bibliography{references}

\begin{thebibliography}{10}
\newcommand{\enquote}[1]{``#1''}

\bibitem{diddams2020optical}
S.~A. Diddams, K.~Vahala, and T.~Udem, \enquote{Optical frequency combs:
  Coherently uniting the electromagnetic spectrum,}
  {\protect\JournalTitle{Science}} \textbf{369}, eaay3676 (2020).

\bibitem{oelker2019demonstration}
E.~Oelker, R.~B. Hutson, C.~J. Kennedy, L.~Sonderhouse, T.~Bothwell, A.~Goban,
  D.~Kedar, C.~Sanner, J.~M. Robinson, G.~E. Marti, D.~G. Matei, T.~Legero,
  M.~Giunta, R.~Holzwarth, F.~Riehle, U.~Sterr, and J.~Ye,
  \enquote{Demonstration of 4.8 $\times$ 10$^{-17}$ stability at 1 s for two
  independent optical clocks,} {\protect\JournalTitle{Nature Photonics}}
  \textbf{13}, 714--719 (2019).

\bibitem{nemitz2016frequency}
N.~Nemitz, T.~Ohkubo, M.~Takamoto, I.~Ushijima, M.~Das, N.~Ohmae, and
  H.~Katori, \enquote{Frequency ratio of {Y}b and {S}r clocks with 5 $\times$
  10$^{- 17}$ uncertainty at 150 seconds averaging time,}
  {\protect\JournalTitle{Nature Photonics}} \textbf{10}, 258--261 (2016).

\bibitem{stern2020direct}
L.~Stern, J.~R. Stone, S.~Kang, D.~C. Cole, M.-G. Suh, C.~Fredrick, Z.~Newman,
  K.~Vahala, J.~Kitching, S.~A. Diddams, and S.~B. Papp, \enquote{Direct {K}err
  frequency comb atomic spectroscopy and stabilization,}
  {\protect\JournalTitle{Science Advances}} \textbf{6}, eaax6230 (2020).

\bibitem{minoshima2000high}
K.~Minoshima and H.~Matsumoto, \enquote{High-accuracy measurement of 240-m
  distance in an optical tunnel by use of a compact femtosecond laser,}
  {\protect\JournalTitle{Applied Optics}} \textbf{39}, 5512--5517 (2000).

\bibitem{ye2004absolute}
J.~Ye, \enquote{Absolute measurement of a long, arbitrary distance to less than
  an optical fringe,} {\protect\JournalTitle{Optics Letters}} \textbf{29},
  1153--1155 (2004).

\bibitem{coddington2009rapid}
I.~Coddington, W.~C. Swann, L.~Nenadovic, and N.~R. Newbury, \enquote{Rapid and
  precise absolute distance measurements at long range,}
  {\protect\JournalTitle{Nature Photonics}} \textbf{3}, 351--356 (2009).

\bibitem{na2020ultrafast}
Y.~Na, C.-G. Jeon, C.~Ahn, M.~Hyun, D.~Kwon, J.~Shin, and J.~Kim,
  \enquote{Ultrafast, sub-nanometre-precision and multifunctional
  time-of-flight detection,} {\protect\JournalTitle{Nature Photonics}}
  \textbf{14}, 355--360 (2020).

\bibitem{riemensberger2020massively}
J.~Riemensberger, A.~Lukashchuk, M.~Karpov, W.~Weng, E.~Lucas, J.~Liu, and
  T.~J. Kippenberg, \enquote{Massively parallel coherent laser ranging using a
  soliton microcomb,} {\protect\JournalTitle{Nature}} \textbf{581}, 164--170
  (2020).

\bibitem{fortier2011generation}
T.~M. Fortier, M.~S. Kirchner, F.~Quinlan, J.~Taylor, J.~C. Bergquist,
  T.~Rosenband, N.~Lemke, A.~Ludlow, Y.~Jiang, C.~W. Oates, and S.~A. Diddams,
  \enquote{Generation of ultrastable microwaves via optical frequency
  division,} {\protect\JournalTitle{Nature Photonics}} \textbf{5}, 425--429
  (2011).

\bibitem{nakamura2020coherent}
T.~Nakamura, J.~Davila-Rodriguez, H.~Leopardi, J.~A. Sherman, T.~M. Fortier,
  X.~Xie, J.~C. Campbell, W.~F. McGrew, X.~Zhang, Y.~S. Hassan, D.~Nicolodi,
  K.~Beloy, S.~A. Diddams, and F.~Quinlan, \enquote{Coherent optical clock
  down-conversion for microwave frequencies with 10$^{-18}$ instability,}
  {\protect\JournalTitle{Science}} \textbf{368}, 889--892 (2020).

\bibitem{liu2020photonic}
J.~Liu, E.~Lucas, A.~S. Raja, J.~He, J.~Riemensberger, R.~N. Wang, M.~Karpov,
  H.~Guo, R.~Bouchand, and T.~J. Kippenberg, \enquote{Photonic microwave
  generation in the {X}-and {K}-band using integrated soliton microcombs,}
  {\protect\JournalTitle{Nature Photonics}} \textbf{14}, 486--491 (2020).

\bibitem{Herr_soliton}
T.~Herr, V.~Brasch, J.~D. Jost, C.~Y. Wang, N.~M. Kondratiev, M.~L. Gorodetsky,
  and T.~J. Kippenberg, \enquote{Temporal solitons in optical microresonators,}
  {\protect\JournalTitle{Nature Photonics}} \textbf{8}, 145--152 (2014).

\bibitem{kippenberg2018dissipative}
T.~J. Kippenberg, A.~L. Gaeta, M.~Lipson, and M.~L. Gorodetsky,
  \enquote{Dissipative {K}err solitons in optical microresonators,}
  {\protect\JournalTitle{Science}} \textbf{361}, eaan8083 (2018).

\bibitem{stern2018battery}
B.~Stern, X.~Ji, Y.~Okawachi, A.~L. Gaeta, and M.~Lipson,
  \enquote{Battery-operated integrated frequency comb generator,}
  {\protect\JournalTitle{Nature}} \textbf{562}, 401--405 (2018).

\bibitem{shen2020integrated}
B.~Shen, L.~Chang, J.~Liu, H.~Wang, Q.-F. Yang, C.~Xiang, R.~N. Wang, J.~He,
  T.~Liu, W.~Xie, J.~Guo, D.~Kinghorn, L.~Wu, Q.-X. Ji, T.~J. Kippenberg,
  K.~Vahala, and J.~E. Bowers, \enquote{Integrated turnkey soliton microcombs,}
  {\protect\JournalTitle{Nature}} \textbf{582}, 365--369 (2020).

\bibitem{trocha2018ultrafast}
P.~Trocha, M.~Karpov, D.~Ganin, M.~H.~P. Pfeiffer, A.~Kordts, S.~Wolf,
  J.~Krockenberger, P.~Marin-Palomo, C.~Weimann, S.~Randel, W.~Freude, T.~J.
  Kippenberg, and C.~Koos, \enquote{Ultrafast optical ranging using
  microresonator soliton frequency combs,} {\protect\JournalTitle{Science}}
  \textbf{359}, 887--891 (2018).

\bibitem{marin2017microresonator}
P.~Marin-Palomo, J.~N. Kemal, M.~Karpov, A.~Kordts, J.~Pfeifle, M.~H. Pfeiffer,
  P.~Trocha, S.~Wolf, V.~Brasch, M.~H. Anderson, R.~Rosenberger, K.~Vijayan,
  W.~Freude, T.~J. Kippenberg, and C.~Koos, \enquote{Microresonator-based
  solitons for massively parallel coherent optical communications,}
  {\protect\JournalTitle{Nature}} \textbf{546}, 274--279 (2017).

\bibitem{zhang2019terahertz}
S.~Zhang, J.~M. Silver, X.~Shang, L.~Del~Bino, N.~M. Ridler, and P.~Del’Haye,
  \enquote{Terahertz wave generation using a soliton microcomb,}
  {\protect\JournalTitle{Optics Express}} \textbf{27}, 35257--35266 (2019).

\bibitem{kuse2019frequency}
N.~Kuse and M.~E. Fermann, \enquote{Frequency-modulated comb lidar,}
  {\protect\JournalTitle{APL Photonics}} \textbf{4}, 106105 (2019).

\bibitem{yu2017microresonator}
M.~Yu, Y.~Okawachi, A.~G. Griffith, M.~Lipson, and A.~L. Gaeta,
  \enquote{Microresonator-based high-resolution gas spectroscopy,}
  {\protect\JournalTitle{Optics Letters}} \textbf{42}, 4442--4445 (2017).

\bibitem{kuse2020continuous}
N.~Kuse, T.~Tetsumoto, G.~Navickaite, M.~Geiselmann, and M.~E. Fermann,
  \enquote{Continuous scanning of a dissipative {K}err-microresonator soliton
  comb for broadband, high-resolution spectroscopy,}
  {\protect\JournalTitle{Optics Letters}} \textbf{45}, 927--930 (2020).

\bibitem{lin2020broadband}
T.~Lin, A.~Dutt, C.~Joshi, C.~T. Phare, Y.~Okawachi, A.~L. Gaeta, and
  M.~Lipson, \enquote{Broadband ultrahigh-resolution chip-scale scanning
  soliton dual-comb spectroscopy,} {\protect\JournalTitle{arXiv preprint
  arXiv:2001.00869}}  (2020).

\bibitem{liu2020monolithic}
J.~Liu, H.~Tian, E.~Lucas, A.~S. Raja, G.~Lihachev, R.~N. Wang, J.~He, T.~Liu,
  M.~H. Anderson, W.~Weng, S.~A. Bhave, and T.~J. Kippenberg,
  \enquote{Monolithic piezoelectric control of soliton microcombs,}
  {\protect\JournalTitle{Nature}} \textbf{583}, 385--390 (2020).

\bibitem{xue2016thermal}
X.~Xue, Y.~Xuan, C.~Wang, P.-H. Wang, Y.~Liu, B.~Niu, D.~E. Leaird, M.~Qi, and
  A.~M. Weiner, \enquote{Thermal tuning of {K}err frequency combs in silicon
  nitride microring resonators,} {\protect\JournalTitle{Optics Express}}
  \textbf{24}, 687--698 (2016).

\bibitem{Gaeta_heater}
C.~Joshi, J.~K. Jang, K.~Luke, X.~Ji, S.~A. Miller, A.~Klenner, Y.~Okawachi,
  M.~Lipson, and A.~L. Gaeta, \enquote{Thermally controlled comb generation and
  soliton modelocking in microresonators,} {\protect\JournalTitle{Opt. Lett.}}
  \textbf{41}, 2565--2568 (2016).

\bibitem{raja2020chip}
A.~S. Raja, J.~Liu, N.~Volet, R.~N. Wang, J.~He, E.~Lucas, R.~Bouchandand,
  P.~Morton, J.~Bowers, and T.~J. Kippenberg, \enquote{Chip-based soliton
  microcomb module using a hybrid semiconductor laser,}
  {\protect\JournalTitle{Optics Express}} \textbf{28}, 2714--2721 (2020).

\bibitem{voloshin2019dynamics}
A.~S. Voloshin, J.~Liu, N.~M. Kondratiev, G.~V. Lihachev, T.~J. Kippenberg, and
  I.~A. Bilenko, \enquote{Dynamics of soliton self{\textendash}injection
  locking in a photonic chip{\textendash}based microresonator,}
  {\protect\JournalTitle{arXiv preprint arXiv:1912.11303}}  (2019).

\bibitem{raja2019electrically}
A.~S. Raja, A.~S. Voloshin, H.~Guo, S.~E. Agafonova, J.~Liu, A.~S.
  Gorodnitskiy, M.~Karpov, N.~G. Pavlov, E.~Lucas, R.~R. Galiev, A.~E.
  Shitikov, J.~D. Jost, M.~L. Gorodetsky, and T.~J. Kippenberg,
  \enquote{Electrically pumped photonic integrated soliton microcomb,}
  {\protect\JournalTitle{Nature Communications}} \textbf{10}, 680 (2019).

\bibitem{Papp_OL_SSB18}
T.~C. Briles, J.~R. Stone, T.~E. Drake, D.~T. Spencer, C.~Fredrick, Q.~Li,
  D.~Westly, B.~R. Ilic, K.~Srinivasan, S.~A. Diddams, and S.~B. Papp,
  \enquote{Interlocking {K}err{\textendash}microresonator frequency combs for
  microwave to optical synthesis,} {\protect\JournalTitle{Opt. Lett.}}
  \textbf{43}, 2933--2936 (2018).

\bibitem{stone2018thermal}
J.~R. Stone, T.~C. Briles, T.~E. Drake, D.~T. Spencer, D.~R. Carlson, S.~A.
  Diddams, and S.~B. Papp, \enquote{Thermal and nonlinear dissipative-soliton
  dynamics in {K}err-microresonator frequency combs,}
  {\protect\JournalTitle{Physical Review Letters}} \textbf{121}, 063902 (2018).

\bibitem{Kippenberg_SiN16}
V.~Brasch, M.~Geiselmann, T.~Herr, G.~Lihachev, M.~H.~P. Pfeiffer, M.~L.
  Gorodetsky, and T.~J. Kippenberg, \enquote{Photonic chip{\textendash}based
  optical frequency comb using soliton cherenkov radiation,}
  {\protect\JournalTitle{Science}} \textbf{351}, 357--360 (2016).

\bibitem{niu2018repetition}
R.~Niu, S.~Wan, S.-M. Sun, T.-G. Ma, H.-J. Chen, W.-Q. Wang, Z.-Z. Lu, W.-F.
  Zhang, G.-C. Guo, C.-L. Zou, and C.~H. Dong, \enquote{Repetition rate tuning
  of soliton in microrod resonators,} {\protect\JournalTitle{arXiv preprint
  arXiv:1809.06490}}  (2018).

\bibitem{zhou2019soliton}
H.~Zhou, Y.~Geng, W.~Cui, S.-W. Huang, Q.~Zhou, K.~Qiu, and C.~W. Wong,
  \enquote{Soliton bursts and deterministic dissipative kerr soliton generation
  in auxiliary-assisted microcavities,} {\protect\JournalTitle{Light: Science
  \& Applications}} \textbf{8}, 50 (2019).

\bibitem{zhang2019sub}
S.~Zhang, J.~M. Silver, L.~Del~Bino, F.~Copie, M.~T. Woodley, G.~N. Ghalanos,
  A.~{\O}. Svela, N.~Moroney, and P.~Del'Haye,
  \enquote{Sub{\textendash}milliwatt{\textendash}level microresonator solitons
  with extended access range using an auxiliary laser,}
  {\protect\JournalTitle{Optica}} \textbf{6}, 206--212 (2019).

\bibitem{lu2019deterministic}
Z.~Lu, W.~Wang, W.~Zhang, S.~T. Chu, B.~E. Little, M.~Liu, L.~Wang, C.-L. Zou,
  C.-H. Dong, B.~Zhao, and W.~Zhao, \enquote{Deterministic generation and
  switching of dissipative {K}err soliton in a thermally controlled
  micro{\textendash}resonator,} {\protect\JournalTitle{AIP Advances}}
  \textbf{9}, 025314 (2019).

\bibitem{nishimoto2020investigation}
K.~Nishimoto, K.~Minoshima, T.~Yasui, and N.~Kuse, \enquote{Investigation of
  the phase noise of a microresonator soliton comb,}
  {\protect\JournalTitle{Optics Express}} \textbf{28}, 19295--19303 (2020).

\end{thebibliography}

\end{document}